# An efficient algorithm for finding all possible input nodes for controlling complex networks

Xizhe Zhang[1,2*], Jianfei Han[2], Weixiong Zhang[3]

[1] (Key Laboratory of Medical Image Computing of Northeastern University, Ministry of Education, China)

[2] (School of Computer Science and Engineering, Northeastern University, Shenyang, Liaoning, China)

[3] (Department of Computer Science and Engineering, Washington University, St Louis, Missouri, USA)

**Abstract:** Understanding structural controllability of a complex network requires to identify a Minimum Input nodes Set (*MIS*) of the network. It has been suggested that finding an *MIS* is equivalent to computing a maximum matching of the network, where the unmatched nodes constitute an MIS. However, maximum matching of a network is often not unique, and finding all *MIS*s may provide deep insights to the controllability of the network. Finding all possible input nodes, which form the union of all *MIS*s, is computationally challenging for large networks. Here we present an efficient enumerative algorithm for the problem. The main idea is to modify a maximum matching algorithm to make it efficient for finding all possible input nodes by computing only one *MIS*. We rigorously proved the correctness of the new algorithm and evaluated its performance on synthetic and large real networks. The experimental results showed that the new algorithm ran several orders of magnitude faster than the existing method on large real networks.

## Introduction

Controlling complex networks [1-3] is of great importance in many applications, such as social networks, biological networks, and technical networks. For example, it has been shown that understanding network controllability can help identify genes responding to viral infection [4] and genes related to cancer [5], as well as assist analyzing metabolic process [6].

A network is said to be controllable if it can be driven from any initial state to a desirable state in finite steps by exerting external control signals on some selected nodes [1], which are called input nodes [7], driver nodes [8] or control nodes [9]. Input nodes of a network can be inferred by finding a maximum matching of a network, which is consisted of the set of maximum edges that do not share nodes [10]. The unmatched nodes related to a maximum matching constitute a Minimum Input nodes Set, or *MIS*. It has been shown that the size of an MIS is closely related to the node degree distribution of the network [8]. Interestingly, the fraction of input nodes is primarily determined by the nodes of low in- and out-degrees [11]. Input nodes have been extensively used in analyzing many real networks, e.g., identifying important proteins in biological networks [4], analyzing interbank networks [12], and increasing the effectiveness of selective modulation of brain networks [13].

Unfortunately, maximum matching is not unique for most networks [14] (Fig. 1). Although the size of these MISs is the same, they may be composed of different input nodes. We call a node in an MIS a *possible input node*. Apparently, all possible input nodes are the union of all MISs. It is essential to find all possible input nodes for understanding the controllability of a complex network. For example, finding all possible input nodes could help understand the roles of nodes in control systems [15], design suitable MISs under different constraints [7], and identify critical genes on signaling pathways [16]. However, finding all possible input nodes by finding all MISs is #*P*-hard [17]. To address this problem, a previous approach first computes a maximum matching and then assess if an unmatched node is a possible input node by removing it to test if its removal may result in a larger maximum matching [15]. The computational complexity of finding a maximum matching is $O(N^{1/2}L)$ and the evaluation process is $O(NL)$ on a network of $N$ nodes and $L$ edges, for a total complexity of $O(NL)$ for the previous method.

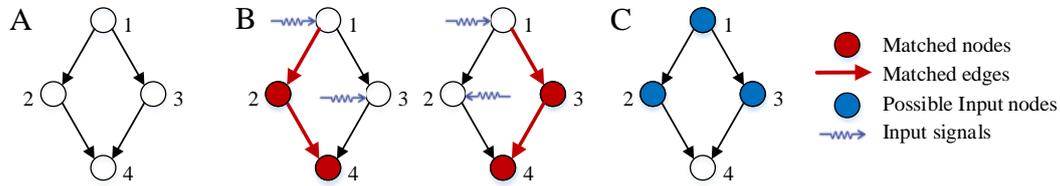

**Figure.1** | An example network with two *MIS*s. A. an example network; B. two *MIS* of the network $D_1=\{1, 3\}$, $D_2=\{1, 2\}$; C. all possible input nodes of the network, which form the union of both *MIS*s.

We developed an efficient algorithm for finding all possible input nodes of a network. We proved that all possible input nodes can be identified by a simple modification to a maximum matching algorithm. The complexity of our algorithm is $O(N^{1/2}L)$, which is the same as the complexity of the maximum matching algorithm. Because our algorithm does not need to evaluate every node of the network, it runs several orders of magnitude faster than the previous method [15] on large networks.

**Method**

Consider a directed network $G(V, E)$ over a set of nodes $V$ and a set of edges $E$. To find an MIS of $G(V, E)$, we first convert the directed network $G(V, E)$ to an equivalent undirected bipartite graph $B(V^{in}, V^{out}, E)$ (Fig. 2A-2B). The bipartite graph is built by splitting the node set $V$ into two node sets $V^{in}$ and $V^{out}$, where a node $n$ in $G$ is converted to two nodes $n^{in}$ and $n^{out}$ in $B$, and nodes $n^{in}$ and $n^{out}$ are, respectively, connected to the in-edges and out-edges of node $n$.

Now consider maximum matching of a bipartite graph. A *matching* is a set of edges that share no common node. A node is called a *matched node* if it is connected to a matching edge, or *unmatched node*, otherwise. A matching with the maximum number of edges is called a *maximum matching*. In an undirected bipartite graph, an *alternate path* is a path whose edges are alternate in and out matching. An *argument path* is an alternate path whose two end nodes are unmatched nodes. Based on the Berge theorem [18], a matching $M^*$ is a maximum matching if there is no augment path in $B(v_1, v_2, E)$ with respect to $M^*$. The input nodes are the unmatched nodes in $V^{in}$ corresponding to a maximum matching of bipartite graph $B(V^{in}, V^{out}, E)$. The unmatched nodes in $V^{in}$ corresponding to any maximum matching form an MIS of $G$.

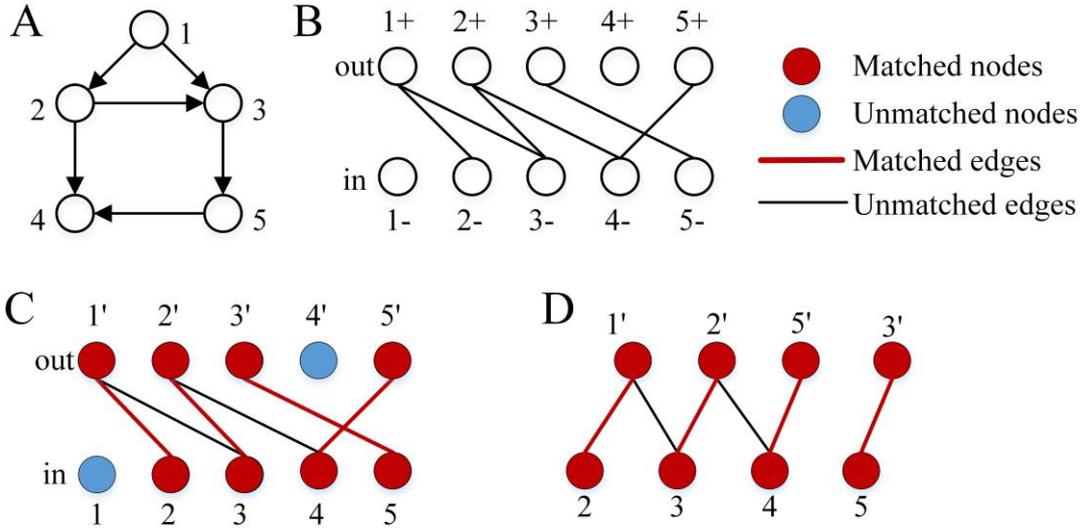

**Figure.2** | An example of a maximum matching of a network. A. a directed network; B. its corresponding bipartite graph; C. A maximum matching of the bipartite graph. An unmatched node in the in-set is an input node; D. Two alternate paths corresponding to the maximum matching in C.

Because maximum matching is not unique for most networks, there may exist many MISs. The union of all MISs is all possible input nodes. We now show that to find all possible input nodes of a network, we only need to compute one maximum matching instead of enumerating all MISs or evaluating all matched nodes as done in [15],

**Theorem 1:** Given a network $G$ and a maximum matching $M$ of $G$, a node $n$ is a possible input node if it satisfies one of the following two conditions:

1. Node $n$ is an input node related to $M$;
2. The in-node $n^{in}$ of the bipartite graph $B$ can be reached from an input node $m^{in}$ related to $M$ through an alternate path $p_{nm}$.

**Proof**: We only need to consider condition 2.

Sufficiency. Suppose that node $n$ satisfies condition 2. Apparently, the length of $p_{nm}$ must be even because both node $n^{in}$ and $m^{in}$ are in the set $V^{in}$ of bipartite graph $B$. Therefore, the alternate path $p_{nm}$ must start with an unmatched edge connected to $m^{in}$ and end with a matched edge connected to node $n^{in}$. Change the types of all edges of $p_{nm}$, i.e., change the matched edges to unmatched and the unmatched edges to matched, then the new path $p'_{nm}$ is still an alternate path. Consequently, we get a new maximum matching $M'$. Clearly, the node $n^{in}$ is not matched by $M'$. Therefore, node $n$ is a possible input node.

Necessity. Let node $n^{in}$ is matched in $M$ and cannot be reached by any input node related to $M$. Suppose that node $n^{in}$ is not matched by a maximum matching $M'$. Node $n^{in}$ must have at least one in-edge because it is matched by $M$. Therefore, there must be an alternate path $p_{nm}$ related to $M'$ which starts with unmatched node $n^{in}$ and end with a matched node $m^{in}$. Now consider the path $p_{nm}$ under the maximum matching $M$. The length of $p_{nm}$ must be even because nodes $n^{in}$ and $m^{in}$ are both in the set $V^{in}$. Therefore, the alternate path $p_{nm}$ must end with an unmatched node $m^{in}$ related to $M$ because $n^{in}$ is matched by $M$. This is in contradict with the fact that $n^{in}$ cannot be reached by any input node related to $M$, which completes the proof.

The significance of the above theorem is that all possible input nodes can be identified by some alternate paths of the input nodes of any given *MIS*. This observation leads to a novel two-step approach to identification of all possible input nodes, i.e., we first compute an *MIS* and then consider its alternate paths. Moreover, these two steps can be combined using a simple modification to the Hopcroft–Karp maximum matching algorithm for undirected graphs [19]. The basic idea of the Hopcroft–Karp algorithm is to iteratively find all argument paths corresponding to the matching *M* at hand, and then to derive a larger matching *M'*. A maximum matching is obtained when no argument path can be founded. The last step of the algorithm is exactly to look for all alternate paths starting from the input nodes of the maximum matching. Therefore, all possible input nodes can be obtained in the last step of Hopcroft–Karp algorithm based on Theorem 1.

The above idea and steps can be formulated in Algorithm *All_Input*(*G*) for finding all possible input nodes in network *G*, which is listed as follows:

*All_Input*(*G*):

1. For a directed network *G*(*V*,*E*), let *B*($V^{out}$, $V^{in}$, *E*) be its corresponding bipartite graph; let the initial matching *M*= null;

2. Find all the alternate paths of all unmatched nodes in $V^{in}$, denoted as *AP* = {$P_1, P_2…P_n$}, and let the nodes of *AP* in $V^{in}$ as candidate results;

3. If *AP* contain argument paths, expand all argument paths and obtain a new matching *M'*; clear all candidate nodes, *M* = *M'*; return to step 2;

4. If *AP* contain no argument path, the candidate nodes are all possible input nodes, and the set of the unmatched nodes is an *MIS* of *G*.

Figure 3 illustrates an example of *All_Input*(*G*) on a small network. The time complexity of the above algorithm is the same as that of the Hopcroft-Karp algorithm, which is $O(N^{1/2}L)$.

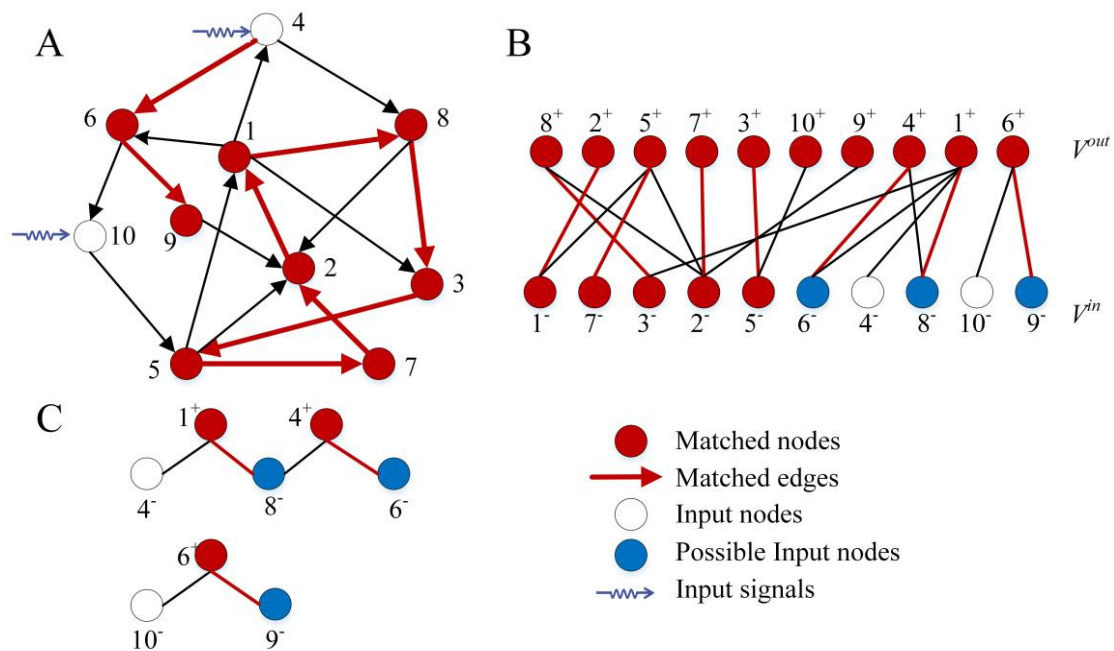

**Figure 3:** An example of the process of algorithm *All_Input*(*G*). **(A)** A sample network and its maximum matching (red edges) **(B)** the corresponding bipartite graph; **(C)** In the last step of the algorithm, we search for alternate paths from unmatched nodes. The nodes on the alternate paths in $V^{in}$ are nodes {8,6,9}, and the input nodes are {4,10}. Therefore, all possible input nodes of the network are {4,6,8,9,10}.

**Result**

To assess the efficiency of the new algorithm, which was coded in JAVA and is available in supplementary information, we compared it with the previous algorithm in [15]. The comparison was done on a Windows 7 workstation with a quad-core Intel i7-3770 processor of 3.9 GHz and 32GB DDR3 1600MHz memory.

We consider 13 synthetic networks, in which the number of nodes $n$ varied from $10^5$ to $5\times 10^6$ and the average degree $<k>$ varied from 6 to 16. Networks were generated with Gephi [20] based on the Scale-Free Network model presents in [21]. As shown in Figure 4, our algorithm significantly outperformed previous algorithm [15]. With small network with $n=10^5$, our algorithm achieved 52x speedup compared to [15]. With larger network with $n=5\times 10^6$, our algorithm achieved 7330x speedup with the execution time is only 3.276 second. Note that the speedup is increased with the average degree $<k>$ (Figure.4A), which indicating that our algorithm has better performance in dense networks. The details of the results are listed in Table.1.

Next, we evaluated the performance of the algorithm in some real networks. These networks are selected based on their diversity of topological structure. These networks include biological networks, social networks, and Internet networks. The size of these networks varied from very small (E.Coli network, 423 nodes) to very large (Amazon network, $4\times 10^6$ nodes). The results shown in Table.1 indicated that our algorithm also significantly outperformed previous algorithm. For very large networks, such as Amazon or Twitter, the results can be obtained within 10 seconds, which are almost $10^4$x speedup compared to the previous algorithm.

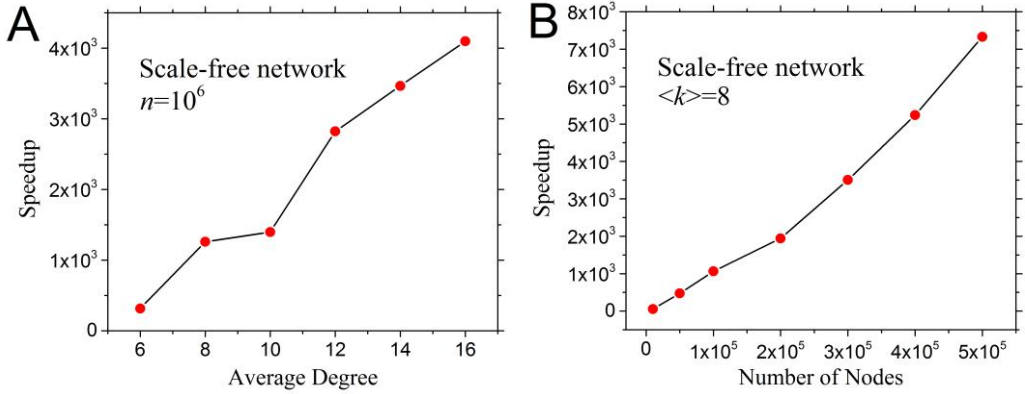

**Figure.4** | Speedup of our algorithm as compared to previous algorithm [15]. A. Speedup as a function of the average degree when $n=10^6$; B. Speedup as a function of the number of nodes when average degree $<k>=8$. The networks are generated based on Scale-Free network model [21] with $r^{in}=r^{out}=3$.

**Table 1**. Comparison of the execution time of some synthetic networks. For each network, we show its average degree $<k>$, number of nodes ($N$) and edges ($L$), destiny of all possible input nodes $N_{pd}$, the execution time of our method $t_{new}$, the execution time of previous method [19] $t_{old}$, and the speedup ratio.

| Network | | L | $n_{pd}$ | $t_{new}(sec)$ | $t_{old}(sec)$ | Speedup |
|---|---|---|---|---|---|---|
| $n=10^6$ | $<k>=6$ | 300000 | 0.444 | 0.343 | 108.1 | 315.1 |
| | $<k>=8$ | 400000 | 0.396 | 0.546 | 687.9 | 1260.1 |

|  | $<k>=10$ | 500000 | 0.124 | 0.858 | 1198.6 | 1396.9 |
|---|---|---|---|---|---|---|
|  | $<k>=12$ | 600000 | 0.039 | 0.826 | 2330.6 | 2821.5 |
|  | $<k>=14$ | 700000 | 0.018 | 0.889 | 3080.3 | 3464.9 |
|  | $<k>=16$ | 800000 | 0.008 | 0.952 | 3900.1 | 4096.8 |
|  | $n=10^5$ | 40000 | 0.332 | 0.047 | 2.5 | 52.1 |
|  | $n=5*10^5$ | 200000 | 0.388 | 0.218 | 102.9 | 472.0 |
|  | $n=10^6$ | 400000 | 0.411 | 0.530 | 562.9 | 1062.2 |
| $<k>=8$ | $n=2*10^6$ | 800000 | 0.397 | 1.435 | 2784.6 | 1940.5 |
|  | $n=3*10^6$ | 1200000 | 0.397 | 2.106 | 7381.4 | 3504.9 |
|  | $n=4*10^6$ | 1600000 | 0.399 | 2.777 | 14550 | 5239.5 |
|  | $n=5*10^6$ | 2000000 | 0.395 | 3.276 | 24012.7 | 7329.9 |

**Table 2**. Comparison of the execution time of some real networks. For each network, we show its type, name, number of nodes (*N*) and edges (*L*), density of all possible input nodes $n_{pd}$, the execution time of our method $t_{new}$, the execution time of previous method [15] $t_{old}$, and the speedup ratio.

| Type | Name | N | L | $n_{pd}$ | $t_{new}$ (*sec*) | $t_{old}$ (*sec*) | *Speedup* |
|---|---|---|---|---|---|---|---|
| **Biological** | E.Coli[22] | 423 | 578 | 0.730 | 0.001 | 0.016 | 16.0 |
|  | TRN-Yeast-1[23] | 4441 | 12873 | 0.999 | 0.015 | 0.062 | 4.1 |
|  | TRN-Yeast-2[24] | 688 | 1079 | 0.920 | 0.001 | 0.015 | 15.0 |
|  | Human PPI [25] | 6339 | 34814 | 0.585 | 0.032 | 1.295 | 40.5 |
| **Trust** | Slashdot0902[26] | 82168 | 948464 | 0.912 | 0.421 | 1568.3 | 3725.2 |
|  | Slashdot0811[26] | 77360 | 905468 | 0.910 | 0.234 | 1388.9 | 5935.5 |
|  | WikiVote[27] | 7115 | 103689 | 0.666 | 0.047 | 2.044 | 43.5 |
|  | SciMet[28] | 3084 | 10416 | 0.661 | 0.015 | 0.187 | 12.5 |
|  | Kohonen[29] | 4470 | 12731 | 0.669 | 0.016 | 0.172 | 10.8 |
| **Internet** | p2p-1[30] | 10876 | 39994 | 0.911 | 0.062 | 2.746 | 44.3 |
|  | p2p-2[30] | 8846 | 31839 | 0.926 | 0.046 | 1.732 | 37.6 |
|  | p2p-3[30] | 8717 | 31525 | 0.933 | 0.031 | 1.700 | 54.8 |
| **Product co-purchasing** | Amazon0302[31] | 262111 | 1234877 | 0.177 | 1.685 | 14119.4 | 8379.5 |
|  | Amazon0312[31] | 400727 | 3200440 | 0.127 | 9.344 | 36696 | 3927.2 |
|  | Amazon0505[31] | 410236 | 3356824 | 0.915 | 7.519 | 45453 | 6045.2 |
|  | Amazon0601[31] | 403394 | 3387388 | 0.053 | 2.886 | 49293 | 17080.1 |
| **Social network** | Twitter [32] | 81306 | 1768149 | 0.800 | 2.230 | 2532.5 | 1135.6 |
|  | Higgs_Twitter[33] | 456626 | 14855842 | 0.297 | 12.589 | 66445.2 | 5278.0 |
|  | UClonline[34] | 1899 | 20296 | 0.819 | 0.016 | 0.296 | 18.5 |
|  | Facebook_348[32] | 572 | 6384 | 0.612 | 0.001 | 0.062 | 62.0 |

**Conclusion**

We developed an efficient algorithm for finding all possible input nodes for controlling complex networks. We proved that once an *MIS* is obtained, all possible input nodes can be efficiently identified without increasing the overall complexity beyond finding the *MIS*. Therefore, our algorithm offers a significant speedup over the previous algorithm on both synthetic networks and many large real networks. Thanks to its efficiency, the new algorithm makes it possible to study controllability of large real-world networks and will have many potential applications in diverse areas.

**Acknowledgments**

This research was supported by the Fundamental Research Funds for the Central Universities of China under grand number N140404011, and the Natural Science Foundation of China under grant number 91546110, and China Scholarship Council under grant number 201606085011, and the Special Program for Applied Research on Super Computation of the NSFC-Guangdong Joint Fund (the second phase).

**Author Contributions**

X.-Z.Z. proved the theorem and designed algorithm. J.-F.H. wrote the code and performed the experiments. X.-Z.Z. and W.-X.Z wrote the paper. All authors reviewed the manuscript.

**Additional Information**

The authors declare no competing financial interests. Correspondence and requests for materials should be addressed to X.-Z.Z. (Email: zhangxizhe@mail.neu.edu.cn).

**Data availability**

All data generated or analyzed during this study are included in this article (and its Supplementary Information files).